# Longitudinal Safety Analysis For Heterogeneous Platoon Of Automated And Human Vehicles

Zi Yang, Xinpeng Wang, Xin Pei*, Shuo Feng,
Dajun Wang, Jianqiang Wang
Tsinghua University
Beijing, China
peixin@tsinghua.edu.cn

S.C. WONG
Department of Civil Engineering
The University of Hong Kong
Hong Kong

*Abstract*—With the recent advancement in environmental sensing, vehicle control and vehicle-infrastructure cooperation technologies, more and more autonomous driving companies start to put their intelligent cars into road test. But in the near future, we will face a heterogeneous traffic with both intelligent connected vehicles and human vehicles. In this paper, we investigated the impacts of four collision avoidance algorithms under different intelligent connected vehicles market penetration rate. A customized simulation platform is built, in which a platoon can be initiated with many key parameters. For every short time interval, the dynamics of vehicles are updated and input in a kinematics model. If a collision occurs, the energy loss is calculated to represent the crash severity. Four collision avoidance algorithms are chosen and compared in terms of the crash rate and severity at different market penetration rate and different locations of the platoon. The results generate interesting debates on the issues of heterogeneous platoon safety.

## I. INTRODUCTION

Rear-end crash accounts for about 81% of total highway crashes in China in 2015 [1] and caused a great amount of physical injuries and economic loss. According to [2], a fully autonomous-driving vehicle equipped with intelligent connected vehicle(ICV) technology will save around $2000 per year by reducing crashes, improving energy efficiency as well as elevating productivity, showing that the ICV seems to be one of the keys to mitigate or even avoid such loss.

One of the great challenge in using ICV to mitigate the collisions is that the road environment is not 100% intelligent and connected. A heterogeneous platoon with both human-driving vehicles(HV) and intelligent connected vehicles will be the main theme over a period of time. While there are rich literature on efficiency and safety of traditional vehicle platoons (e.g. [3]–[6]), studies relates to the heterogeneous platoon safety problems are very limited.

Most of these heterogeneous platoon safety studies are simulation studies since there are not much field data available. One of the early studies conducted by Carbaugh, Godbole, and Sengupta [7] compares safety of automated and manual highway systems with respect to rear-end collision frequency and severity. They found that automated driving is safer than the most alert manual drivers, at similar speeds and capacities. They also present a detailed safety-capacity trade-off study for four different automated highway system concepts that differ in their information structure and separation policy. But the platoons they are looking at are still two-vehicle systems, which is quite different from the real cases, especially those with large traffic volumes.

Kikuchi, Uno and Tanaka examined how the presence of vehicles equipped with the adaptive cruise control system (ACCS) affects stability and safety of a flow consisting of both ACCS and non-ACCS vehicles[8]. They found that, generally, shorter perception-reaction time of the ACCS vehicles can shorten the process of achieving stability, and also can promote safety to both the ACCS and the non-ACCS vehicles under congested conditions. But their focus is on the different ACCS vehicle's placement in the platoon at the same market penetration rate (MPR).

Segata and Cigno conducted a simulation study of an emergency braking (EB) application accomplished by embedding mobility, cars' dynamic, and drivers' behavior models into a detailed networking simulator[9]. They found that even with very low market penetration rate, the introduction of ICV will not only significantly reduce the number of rear-end collisions, but also improve the safety of unconnected vehicles. But their study only considers cooperation in terms of vehicle-to-vehicle communication, not cooperative control.

In this paper, we present a simulation study focusing on the comparison of rear-end collision avoidance (CA) algorithms at different MPR in a heterogeneous platoon with both HVs and ICVs. The rest of the paper is organized as follows: the second section describes the design of the simulation, including the initialization, CA algorithms, vehicle kinematic model, and the crash simulation; the third section present simulation results and discuss CA algorithms' performances at different MPR; the final section concludes the paper.

## II. SIMULATION DESIGN

The simulation platform is designed and constructed in MATLAB. Compared to other simulation softwires such as NS-3(NS-2), IDM, Vissim, and Prescan, a customized platform in MATLAB is more flexible in both data processing

This study is supported by the NSFC grant (71671100) and Beijing Municipal Science and Technology Program(20171090204）．

and algorithm application to cope with real-world scenes and complex collision avoid algorithms. The simulation flow is shown in Fig. 1. Before each iteration, the parameters are set in the first step. For the second step, the deceleration of each vehicle is calculated with the vehicle dynamics and collision avoidance algorithms. For the third step, collision occurrence is judged with the refreshed vehicle dynamics. If there is no collision, go back to step two and refresh the vehicle dynamics again. If there is a collision, a collision is simulated and the results are recorded.

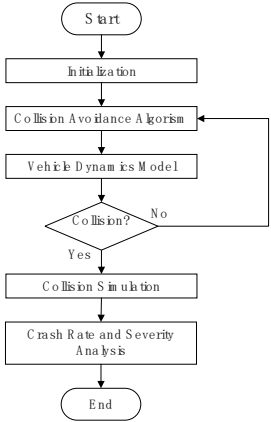

Figure 1 Simulation schematic diagram

### A. Initialization

The simulation platform can customize the number of vehicles in the platoon. For easy horizontal comparison, we use a 11-car fleet to perform the simulation with 1 lead car and 10 following cars. In each iteration, the lead car starts to decelerate at maximum deceleration. The following cars behave according to some collision avoidance algorithms. When all the cars have stopped, the iteration ends. All the vehicle statuses are updated at 10Hz, which is a common communication refresh rate for intelligent connected vehicles. According to single pilot trials, each simulation lasts about 20s.

The cars in the platoon can be either HV or ICV. The ratio of ICV, which is represented as MPR, can be set from 0 to 1 with 0.1 as the interval. The distribution of ICV in the platoon is randomly generated. The masses of cars are evenly distributed from 900kg to 2500kg. The length of vehicle ranges from 3.5m-5.5m and is set in proportion to the mass. The maximum deceleration (m/s$^2$) is set according to [10], which follows N(5.5, 0.6). The following distance is set according to [11], which is evenly distributed between 100km/h and 110km/h. The headway (in meters) follows N(2.0. 0.3).

### B. Collision avoidance algorithms

When the parameters are set, the simulation starts and the dynamics of vehicles starts to update. Vehicles behave according to their respective Collision Avoidance(CA) algorithms.

For HVs, we adopt a basic form of stimulus-response linear car-following model in [12], where the acceleration can be presented as:

$$a_{des}^i(t) = \alpha[v_{i-1}(t - \Delta t) - v_i(t - \Delta t)] \quad (1)$$

Where α is the sensitivity of the driver's reaction to stimulus, and the desired acceleration rate of vehicle *i* at moment *t* is proportional to the relative speed between vehicle *i* and its preceding vehicle Δt ago.

Both sensitivity and driver reaction time for this model come from real-world driving experiment in [13]. The reaction time (in seconds) of drivers are found to follow N(1.1, 0.22). But a problem is that the stimulus in the field test is usually tiny, while the stimulation in real emergency situation could be intense. Thus, we choose the distribution of maximum sensitivity of each driver as the distribution of driver sensitivity in this study, which follows N(0.85, 0.2). Pilot trials show that this parameter generates reasonable results.

For ICVs, we choose four typical CA algorithms:

*1) Collision avoidance based on direct braking (DB)*

The desired acceleration will reach the maximum deceleration instantly and keep until the vehicle come to a complete stop, which is represented as:

$$a_{des}^i(t) = Dec_{max}^i \quad (2)$$

*2) Collision avoidance based on safe distance (SD)*

This algorithm is based on [9]. First, a safe distance is defined:

$$s_{safe}^i(t) = T_{THW} * v_i(t) + \epsilon \quad (3)$$

$T_{THW}$ is time headway, which is set to 1s according to [14]. $\epsilon$ is set to 1m. Then, the desired acceleration of vehicle *i* at time *t* is calculated as:

$$a_{des}^i(t) = \frac{v_{i+1}^2(t) - v_i^2(t)}{2(s_{actual}(t) - s_{safe}(t))} \quad (4)$$

The significance of this acceleration is the minimum deceleration to achieve the same speed before the following distance reaches the safe distance.

*3) Cooperative collision avoidance algorithm based on sliding mode control (SMC)*

This algorithm comes from the cooperative adapted cruise control (CACC). Although it is designed for car-following scenario in steady state, the sudden speed change of the leading vehicle can be regarded as a disturbance in the original model. First, we define a sliding surface:

$$S_i = \dot{\epsilon}_i + \frac{\omega_n}{\xi + \sqrt{\xi^2 - 1}} \frac{1}{1 - C} \epsilon_i + \frac{C}{1 - C}(V_i - V_l) \quad (5)$$

Where $V_i$ is the speed of vehicle I, and $V_l$ is the speed of the leading vehicle. *C* is a constant between 0 and 1 that determines the weighs of the lead car and the preceding car in ego car decision-making. In this study, C is set to 0.7 according to pilot trials. $\omega_n$ is the controller bandwidths, which is set to 0.8 rad/s according to pilot trials. $\xi$ is the damping parameter and is set to 1 according to [15].

We assume:

$$\dot{S}_i = -\lambda S_i \text{ with } \lambda = \omega_n(\xi + \sqrt{\xi^2 - 1}) \quad (6)$$

Thus, the desired acceleration of ego vehicle is calculated as:

$$a_{des}^i = (1-C)a_{i-1} + Ca_1 - \left(2\xi - C(\xi + \sqrt{\xi^2-1})\right)\omega_n \dot{\epsilon}_i - (\xi + \sqrt{\xi^2-1})\omega_n C(V_i - V_1) - \omega_n^2 \epsilon_i \quad (7)$$

*4) Cooperative collision avoidance algorithm based on minimum total relative kinetic energy density (TKED)*

According to [16], if all vehicles are ICVs, i.e. the MPR is 1, and there is a central controller that coordinate the braking force of every vehicle in real-time, we could define a metric to quantify the severity of a collision:

$$F = \frac{m_2}{2S} \cdot (v_1 - v_2)^2, \quad v_1 \leq v_2 \quad (8)$$

*F* can be viewed as the density of relative kinetic energy. According to *F*, we can reach an object function for optimizing the braking behavior of the whole platoon:

$$J(t) = \sum_{i=1}^{n-1} F_{i,i+1}(t) = \frac{1}{2}\sum_{i=1}^{n-1} \frac{m_{i+1}}{x_i(t)-x_{i+1}(t)-L_i} \cdot [v_i(t) - v_{i+1}(t)]^2 \quad (9)$$

However, if *N* out of *M* vehicles are ICVs, then we can define a subset of {1, 2,…M} which records the position of all ICVs:

$$\boldsymbol{G} = \{g_1 \cdots, g_m\}, \quad \boldsymbol{G} \subset \{1, \cdots N\} \quad (10)$$

Assume that every ICV can detect the kinetic state of if preceding HV and its following HV, the set of vehicles that will be considered in this optimization problem is:

$$\boldsymbol{C} = \boldsymbol{G} \cup \{g_i - 1\} \cup \{g_j + 1\}, \quad g_i > 1, g_j < N, \quad g_i, g_j \in \boldsymbol{G} \quad (11)$$

It means all the ICVs and the HVs beside the ICVs will be taken into consideration. Thus, we can build a model predictive control (MPC) model and solve the desired acceleration of each vehicle at each time step.

*C. Vehicle Kinematics Model*

Current state of the vehicle is updated real-time according to the current desired acceleration as well as the state of the vehicle at the last moment. The model implemented is a 1st-order kinetic model, which can be written as:

$$\begin{cases} \dot{x}_i(t) = v_i(t) \\ \dot{v}_i(t) = a_i(t) \\ \dot{a}_i(t) = \frac{1}{\tau}(a_{des}^i(t) - a_i(t)) \end{cases} \quad (12)$$

τ represents the lag between real and desired acceleration, which takes a typical value of 0.5 according to [15]. But since the time in this study is discrete with *Δt=0.1s*, (12) is actually represented as:

$$\begin{cases} x_i(k+1) = v_i(k)\Delta t + x_i(k) \\ v_i(k+1) = a_i(k)\Delta t + v_i(k) \\ a_i(k+1) = \frac{\tau - \Delta t}{\tau} a_i(k) + \frac{\Delta t}{\tau} a_{des}^i(k) \end{cases} \quad (13)$$

*D. Crash Simulation*

In this study, both collision and its severity are considered. A collision occurs when the gap between 2 vehicles is smaller than 0.05m. When a collision happens, the speed of involving vehicles after the collision is calculated based on the mass of the two, the velocity of the two before collision and a predefined coefficient of restitution (COR), which is initially set to 0 according to [9]. Thus, the velocity of involved vehicles are calculated as:

$$v'_{i-1} = \frac{(m_{i-1}-C_R m_i)v_{i-1}+(1+C_R)m_i v_i}{m_{i-1}+m_i} \quad (14)$$

$$v'_i = \frac{(m_i - C_R m_{i-1})v_i + (1+C_R)m_{i-1}v_{i-1}}{m_{i-1}+m_i} \quad (15)$$

Thus, we can also calculate the kinetic energy loss of such a crash, which is represented as:

$$E_{loss}(i-1,i) = \frac{1}{2}(m_{i-1}v_{i-1}^2 + m_i v_i^2 - m_{i-1}v'^2_{i-1} - m_i v'^2_i) \quad (16)$$

All eligible crashes are recorded, but we made two simplifications for analysis convenience: 1) Only the first crash between two tandem vehicles will be recorded. 2) If a vehicle has been hit by its follower, its contacts it may have with its predecessor are ignored.

III. SIMULATION RESULTS

*A. Market Penetration Rate and Rear-end Collisions*

We conducted 11 groups of simulation on MPRs from 0 to 1 with 0.1 as the interval. Here the MPR is calculated using the 10 following cars excluding the lead car. For all CA algorithms except TKED, we run 1000 iterations for each MPR. For TKED algorithm, we only run 100 iterations since the algorithm is rather time-consuming. For each algorithm, we plotted the amount of crashes at each position of the platoon, the average crash rate and the average kinetic energy loss per crash.

The results using SD algorithm are shown in Fig. 3, where car 1 is the lead car at the front of the platoon. It can be seen that the number of crashes decreases from front to the back. This is accorded with the conclusion in previous studies based on linear car-following model that the danger at the lead vehicle is actually attenuating as it propagates through the platoon even though no ICV is present. Now we can extend this conclusion with the participation of ICVs at any MPR. As MPR goes up, i.e. there are more ICVs, the number of crashes at every position of the platoon reduced. This indicates that higher MPR generally gives rise to lower crash probability and better safety for every position in the platoon, especially at lower MPR. For the two figures on the lower part, it can be seen that as the MPR increases, the average crash rate as well as total energy loss all go down. The curve is steeper before the MPR reaches 0.5, and then becomes flatter down until the MPR reaches 1. The crash rate reduces from 0.44 to 0.02 and the energy loss also reduced several orders.

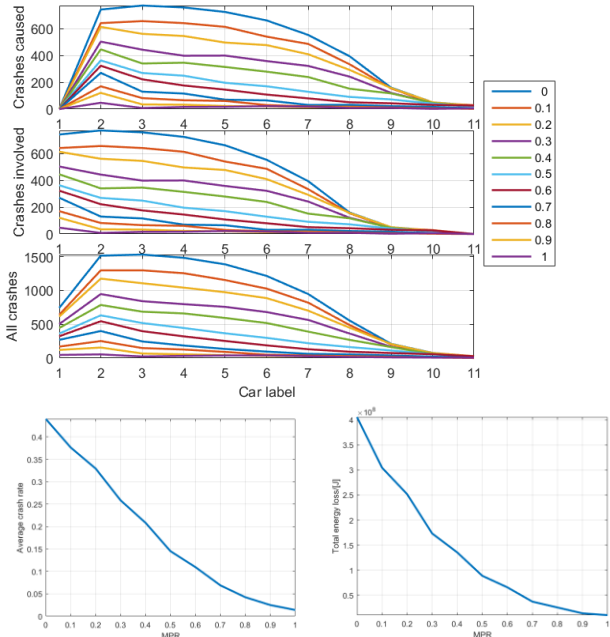

Figure 2 Simulation results using SD

The results using DB algorithm are shown in Fig. 4. It could be seen that when MPR is 0, the curves are quite like those using SD algorithm. But for other MPRs, the vehicle at the back of platoon has similar number of crashes with vehicles at the front. The attenuation of danger through the platoon nearly disappeared. The introduction of ICV actually exacerbates the danger at the back. It can be seen that as the MPR goes from 0 to 0.1, the crash and total energy both increases. Little number of ICVs on the road will make it more dangerous. But they will decrease almost linearly as the MPR keeps going up, especially when MPR reaches 0.3 or higher.

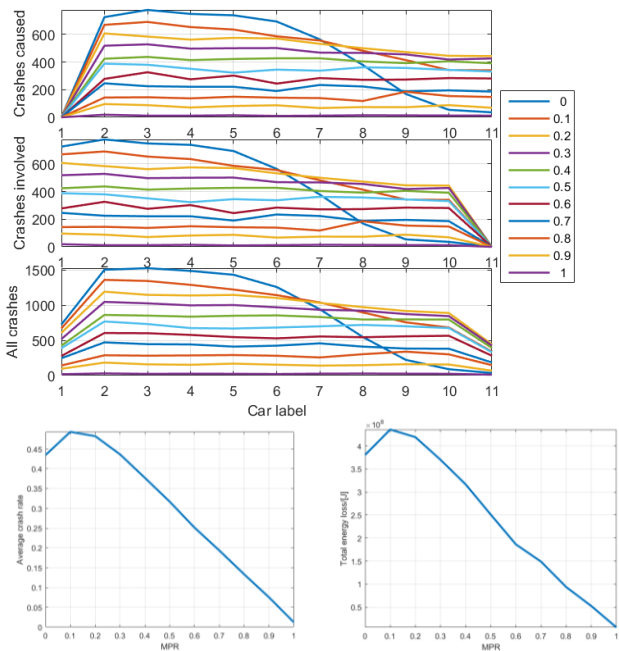

Figure 3 Simulation results using DB

The results using SMC algorithm is shown in Fig. 5. The crash curves are quite like those using DB algorithm. But the crash rate and the total energy loss curves are different. The crash and total energy both decrease monotonically as the MPR keeps going up from 0.

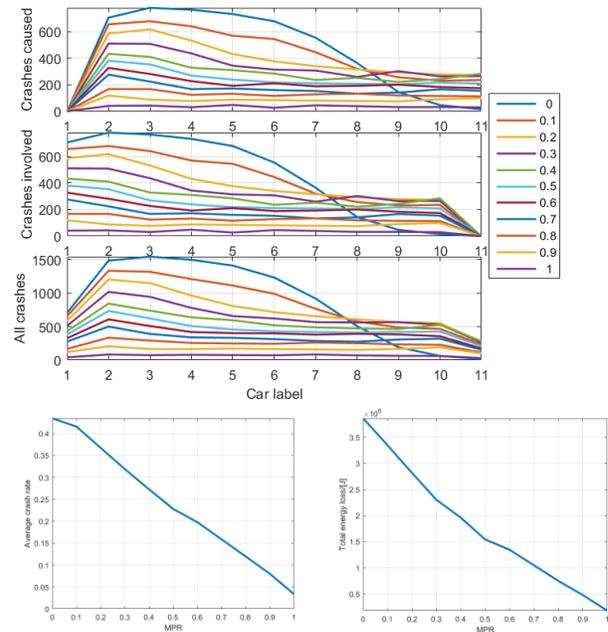

Figure 4 Simulation results using SMC

The results using TKED algorithm is shown in Fig. 6. It can be seen that for all MPRs, the number of crashes decreases from front to back, similar to the results using SD. As MPR increases, the crash rate and total energy loss both go down.

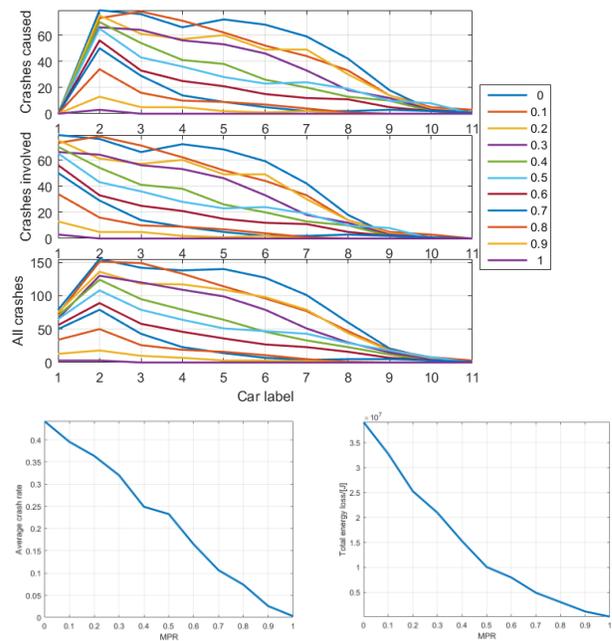

Figure 5 Simulation results using TKED

### B. Comparison of Results using Different CA Algorithms

We have conducted simulations using SD, DB, SMC, and TKED respectively. In this section, their results are compared under different MPR context.

First, the average crash rates using SD, DB, SMC, and TKED are shown in Fig. 7, and the energy losses are shown in Fig.8. It can be seen that for all algorithms except for DB, the average crash rate decreases as the MPR grows. SD have the strongest efficacy for eliminating crashes at low MPR (<80%). While TKED ranked 2nd for low MPR, it can theoretically eradicate all crashes when the MPR reaches 1. Similar trend can be found using energy loss rather than crash rate.

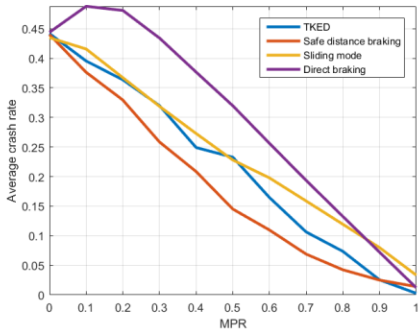
Figure 6 Average crash rate using using SD, DB, SMC, and TKED

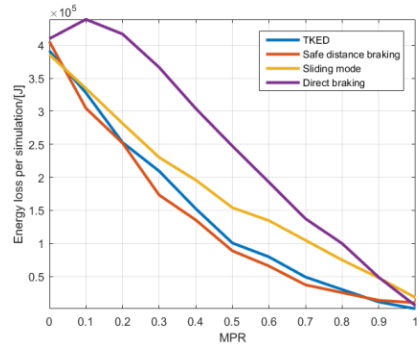
Figure 7 Energy loss using using SD, DB, SMC, and TKED

Then, the histograms of energy loss per crash under three typical MPRs (Low: 0.2, Median: 0.5, and High: 0.8) are shown below in Fig. 9, 10 and 11. It can be seen that the average energy loss is the smallest when using TKED algorithm, i.e. the least severity of crashes. But since this algorithm is minimizing the energy loss, it is not fair to make comparison to results using other algorithms. The energy loss is highest when DB is used, i.e. the highest severity of crashes. When three figures are compared, it could be found that when MPR is higher, the average energy loss from all four CA algorithms are lower. It means when there are more ICVs, the severity of crashes will generally be lower.

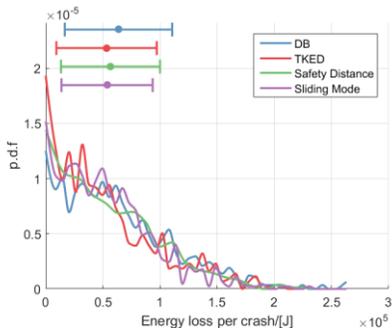
Figure 8 Energy loss per crash when MPR = 0.2

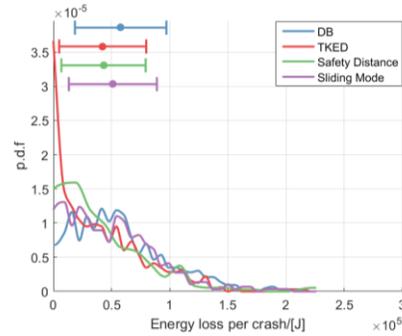
Figure 9 Energy loss per crash when MPR = 0.5

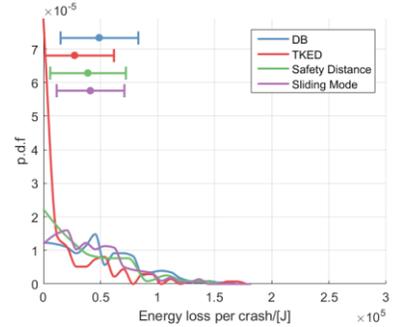
Figure 10 Energy loss per crash when MPR = 0.8

Finally, the locations of crashes in the platoon under three typical MPRs (Low: 0.2, Median: 0.5, and High: 0.8) are shown in Figure 12, 13 and 14.

It can be seen that under low MPR, the front cars have quite similar crash numbers. But start from the car 6, DB has the highest number of crashes. At the back of the platoon (car 9, 10, and 11) SD and TKED has almost no crashes.

When MPR reaches median value, results are a little different. TKED algorithm has higher crash numbers at the front of the platoon (car 2, 3, and 4). SD has the lowest number of crashes at almost every location of the platoon.

When MPR is high, TKED has the highest number of crashes between the lead car and the second car, but as car number goes up, the number of crashes reduced rapidly. SD has the lowest number of crashes at every location of the platoon, especially at the middle and the back of the platoon.

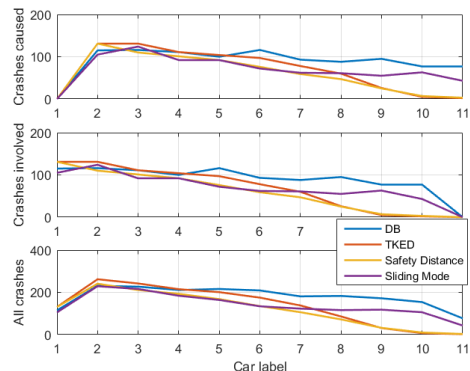
Figure 11 Crash location in the platoon when MPR = 0.2

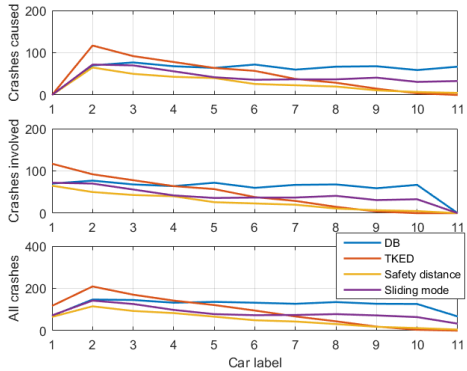

Figure 12 Crash location in the platoon when MPR = 0.5

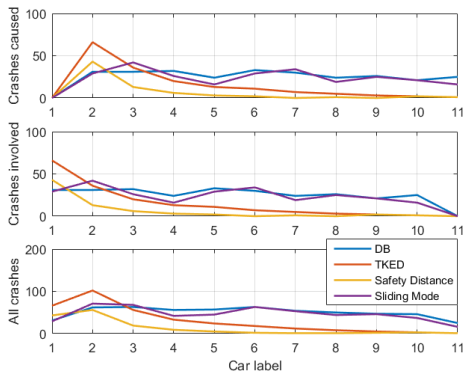

Figure 13 Crash location in the platoon when MPR = 0.8

To sum up, when using DB or SMC algorithms, vehicles at the back of the platoon still have a lot of crashes. But SD and TKED showed a clear attenuation of danger through the platoon, thus the vehicles at the back of the platoon is much safer compared to the vehicles using DB or SMC.

## IV. CONCLUSION AND FUTURE WORKS

It is believed that heterogeneous platoon with both human vehicles and intelligent connected vehicles will go over for a period of time, during which there are many safety problems worth concerning. In this paper, we presented a simulation approach to examine the impacts of four collision avoidance algorithms under different ICV market penetration rate.

As the MPR of ICV going up, the crash rate and the crash severity are reduced for all four algorithms. One special case is when MPR goes from 0 to 0.1, i.e. there are little number of ICVs on the road, DB algorithm increases them. Four algorithms behave differently when attenuating the danger from the lead car through the platoon. The back of the platoon is safer than the front of the platoon, but the attenuation efficacy at different locations of the platoon are different. Generally, SD and TKED algorithms turns to produce the lowest crash rate and crash severity compare to DB or SMC algorithms, especially at median and high MPRs. Even at low MPRs, the SD and TKED will produce safer results in the middle and the back of the platoon.

These findings are hoped to be helpful when forming controlling strategies and enlightening better algorithms for collision avoidance. As we can see, there are still much future works to be done. For example, we believe that the locations of the ICVs also matters in the car-following analysis; the vehicles are not necessarily small cars, but also could be heavy trucks; moreover, in connected vehicle systems, the communication quality is also a key factor. We believe with modifications, our simulation platform will be able to investigate more in the field of heterogeneous platoon safety.